\documentclass[pra,onecolumn,notitlepage,showpacs,showkeys,nofootinbib]{revtex4-1}

\usepackage[english]{babel}
\usepackage[T1]{fontenc}
\usepackage[utf8]{inputenc}

\usepackage{amssymb,amsfonts,amsmath}
\usepackage{braket}
\usepackage{graphicx}
\usepackage{hyperref}
\usepackage{color}
\usepackage{enumerate}

\newcommand{\Uni}[0]{S_{(q,s)}}
\newcommand{\Ry}[0]{S^\mathrm{R}_q}
\newcommand{\Ts}[0]{S^\mathrm{T}_q}
\newcommand{\Tr}[0]{\operatorname{Tr}}
\newcommand{\rank}[0]{\operatorname{rank}}
\newcommand{\Dist}[0]{D^\Pi_{(q,s)}}
\newcommand{\Pty}[0]{\left(\Tr (\rho^{AB})^q \right)^s}
\newcommand{\ketbra}[2]{\ket{#1}\bra{#2}}

\newcommand{\HAB}{\mathcal{H}^{N^{AB}}}
\newcommand{\HA}{\mathcal{H}^{N^A}}
\newcommand{\HB}{\mathcal{H}^{N^B}}

\newcommand{\E}[0]{\mathcal{E}}

\newcommand{\Pc}[1]{\pi_{(q,s)}^{#1}\,} 

\newcommand{\MDA}[0]{D_{(q,s)}^{\Pi^A}(\rho^{AB})}
\newcommand{\MDB}[0]{D_{(q,s)}^{\Pi^B}(\rho^{AB})}
\newcommand{\MDk}[0]{D_{(q,s)}^{\Pi^{K}}(\rho^{AB})}
\newcommand{\MDAB}[0]{D_{(q,s)}^{\Pi^{AB}}(\rho^{AB})}
\newcommand{\mMDA}[0]{D_{(q,s)}^{A}(\rho^{AB})}
\newcommand{\mMDB}[0]{D_{(q,s)}^{B}(\rho^{AB})}
\newcommand{\mMDAB}[0]{D_{(q,s)}^{AB}(\rho^{AB})}
\newcommand{\mMDk}[0]{D_{(q,s)}^{K}(\rho^{AB})}

\newcommand{\MDAc}[1]{D_{(q,s)}^{\Pi^A}(#1)}
\newcommand{\MDBc}[1]{D_{(q,s)}^{\Pi^B}(#1)}

\newcommand{\MDAcc}[2]{D_{(q,s)}^{\Pi^{A}_{#1}}(#2)}
\newcommand{\MDBcc}[2]{D_{(q,s)}^{\Pi^{B}_{#1}}(#2)}
\newcommand{\MDABcc}[2]{D_{(q,s)}^{\Pi^{AB}_{#1}}(#2)}

\begin{document}

\title{Unified  entropic  measures  of  quantum correlations  induced  by  local
  measurements}

\author{G.M.    Bosyk$^{1}$,     G.    Bellomo$^{1}$,    S.    Zozor$^{2}$,
  M. Portesi$^1$ and P.W. Lamberti$^{3}$}
\address{$^1$~Instituto   de   F\'{\i}sica  La   Plata   (IFLP),  CONICET,   and
  Departamento  de  F\'{\i}sica,   Facultad  de  Ciencias  Exactas,  Universidad
  Nacional de La Plata, C.C.~67, 1900 La Plata, Argentina}

\address{$^2$~Laboratoire  Grenoblois  d'Image,  Parole, Signal  et  Automatique
  (GIPSA-Lab, CNRS),  11 rue des Math\'ematiques, 38402  Saint Martin d'H\`eres,
  France}

\address{$^3$~Facultad  de   Matem\'atica,  Astronom\'ia  y   F\'isica  (FaMAF),
  Universidad Nacional  de C\'ordoba, and  CONICET, Avenida Medina  Allende S/N,
  Ciudad Universitaria, X5000HUA, C\'ordoba, Argentina}
%

\begin{abstract}
  We  introduce quantum  correlations measures  based on  the minimal  change in
  unified entropies  induced by local rank-one  projective measurements, divided
  by a factor that  depends on the generalized purity of the  system in the case
  of  non-additive  entropies.   In this  way,  we  overcome  the issue  of  the
  artificial increasing of  the value of quantum correlations  measures based on
  non-additive entropies when an uncorrelated ancilla is appended to the system,
  without changing the computability  of our entropic correlations measures with
  respect to the previous ones.  Moreover, we recover as limiting cases the
  quantum  correlations measures  based  on von  Neumann  and R\'enyi  entropies
  (i.e.,  additive  entropies), for  which  the  adjustment factor  becomes
  trivial.   In   addition,  we   distinguish  between  total   and  semiquantum
  correlations and  obtain some  relations between them.  Finally, we  obtain analytical expressions of
  the entropic correlations measures for typical quantum bipartite systems.
\end{abstract}

%

\maketitle

\section{Introduction}
\label{sec:Introduction}

Quantum correlations  lie at the heart  of the difference  between classical and
quantum worlds.  There  are at least two paradigms to  address this issue beyond
the  usual  entangled-separable  distinction~\cite{Werner1989}.   For  instance,
steering  correlations  have  recently  been  formulated in  a  operational  way
in~\cite{Wiseman2007}, although their origins can  be found in the seminal works
by      Einstein,      Podolski      and      Rosen~\cite{Einstein1935}      and
Schrödinger~\cite{Schrodinger1935}.   These  correlations  intermediate  between
entanglement   and  nonlocality~\cite{Bell1964}   (i.e.,  a   violation  of
Bell inequalities).  On  the  other hand,  it  possible to  identify
quantum  correlations even  in  separable states.   This  has been  firstly
observed by Ollivier and Zurek and Henderson and Vedral, who derived the quantum
discord    as   a    signature    of   quantum    correlations   in    bipartite
systems~\cite{Ollivier2001, Henderson2001}.  The  original definition of discord
relies  on  the  difference  between  two extensions  of  the  classical  mutual
information  to the  quantum  case.   A generalization  of  discord using  other
entropic    forms     by    a     direct    replacement    of     von    Neumman
entropy~\cite{vonNeumann1927}       by       general       entropies,       like
R\'enyi~\cite{Renyi1961}   or  Tsallis~\cite{Tsallis1988}   ones,   as  proposed
in~\cite{Hou2014},   fails  as   it  has   been   shown  in~\cite{Jurkowski2013,
  Bellomo2014}.

Here, we aim  to obtain quantum correlations measures  by using general entropic
forms,    namely   $(q,s)$-entropies   (or    unified   entropies)~\cite{Hu2006,
  Rastegin2011}.   To  avoid  the difficulty  discussed  in~\cite{Jurkowski2013,
  Bellomo2014},  we follow  an  alternative  approach inspired  by  the work  of
Luo~\cite{Luo2008}.   We propose  as quantum  correlations measures  the minimal
change in unified entropies induced  by a local rank-one measurement, divided by
a factor that depends on the  generalized purity only in the case of nonadditive
entropies    (this    adjusting   factor    becomes    trivial   for    additive
entropies). Several  quantum correlations measures discussed  in the literature,
like~\cite{Rajagopal2002, Horodecki2005,  Luo2008, Modi2010, Dakic2010, Luo2010,
  Rossignoli2010,    Rossignoli2011,    Costa2013,   Tufarelli2013,    Zhao2013,
  Canosa2015}, among others, are particular  cases of (or close to) our proposal
(see~\cite{Modi2012} for a recent  review of quantum correlations).  Indeed, the
case of  trace form entropies~\cite{Wehrl1978}, which  are nonadditive entropies
(except  the   von  Neumann  case),  has   been  dealt  in~\cite{Rossignoli2010,
  Rossignoli2011, Costa2013} and deserves  a particular mention.  These entropic
quantum correlations measures artificially increase when an uncorrelated ancilla
is appended to  the system (the geometric discord~\cite{Dakic2010}  has the same
issue, as  it has  been pointed out  in~\cite{Piani2012}). The  nonadditivity of
trace form entropies is the cause of this problem.  We solve this in the case of
$(q,s)$-entropies by introducing a  generalized purity factor, similarly to what
has  been  done  with  the  geometric  discord,  that  is  dividing  it  by  the
purity~\cite{Tufarelli2013}.    In   this   way,   we   obtain   a   family   of
$(q,s)$-entropic measures  of quantum correlations that are  invariant under the
addition  of  an  uncorrelated  ancilla,  both  in the  cases  of  additive  and
non-additive entropies.  In addition,  the computability of our entropic quantum
correlations  measures remain equal  to the  previous ones~\cite{Rossignoli2010,
  Rossignoli2011, Costa2013}, since the adjustment factor is simply the trace of
a power of the density operator.

The outline of this work is as  follows. Our proposal and main results are given
in Sec.~\ref{sec:QC}.   In~\ref{subsec:entropies}, we review  the notion
and some  properties of $(q,s)$-entropies  and majorization, and we  introduce a
family  of entropic  measures of  disturbance due  to a  projective measurement.
In~\ref{subsec:qcm},  we  introduce the  general  entropic quantum  correlations
measures  by  quantifying disturbances  due  to  local projective  measurements,
distinguishing between total and  semiquantum correlations.  Besides, we provide
basic properties that justify our proposal.  In~\ref{subsec:lowerbound}, we find
a  lower bound  of the  entropic quantum  correlations in  terms  of generalized
entanglement  entropy.   In~\ref{subsec:total_semi_qcm},  we  establish  some
interesting  relationships between  total  and semiquantum  measures.  Then,  in
Sec.~\ref{sec:examples} we present some  typical examples where we apply
our   correlations   measures.   Finally,   some   conclusions   are  drawn   in
Sec.~\ref{sec:conclusion}.

\section{Entropic measures of quantum correlations}
\label{sec:QC}

\subsection{Unified entropies, majorization and $(q,s)$-disturbances}
\label{subsec:entropies}

Let  a quantum system  be described  by a  density operator  $\rho$, that  is, a
trace-one positive  semidefinite operator  acting on an  $N$-dimensional Hilbert
space, $\mathcal{H}^N$.  The quantum  unified $(q,s)$-entropies of the state are
defined as~\cite{Hu2006, Rastegin2011},
\begin{equation} \label{eq:Unient}
\Uni(\rho) = \frac{\left( \Tr \rho^q \right)^s - 1}{(1-q)s},
\end{equation}
for  entropic indexes $q  > 0$,  $q \neq  1$ and  $s \neq  0$.  Notice  that the
quantum Tsallis entropies~\cite{Tsallis1988} are obtained for $s=1$,
\begin{equation} \label{eq:Tsallisent}
S_{(q,1)} \equiv \Ts(\rho) = \frac{\Tr \rho^q -1}{1-q},
\end{equation}
being an interesting  case $q=2$, $S_{(2,1)} \equiv S_2(\rho)  = 1- \Tr \rho^2$,
which is directly  related to the purity  of the state.  On the  other hand, von
Neumman  entropy~\cite{vonNeumann1927}  is recovered  in  the  limiting case  $q
\rightarrow 1$,
\begin{equation}
S_{(1,s)} \equiv S(\rho)= - \Tr \rho \ln \rho, \label{eq:vNent}
\end{equation}
whereas R\'enyi entropies~\cite{Renyi1961} are recovered in the limiting case $s
\rightarrow 0$,
\begin{equation}
S_{(q,0)} \equiv \Ry(\rho) = \frac{\log \Tr \rho^q}{1-q}. \label{eq:Ryent}
\end{equation}

A feature of $(q,s)$-entropies is their nonadditive character~\cite{Hu2006},
which is  reflected in the sum  rule for product states  $\rho^A \otimes \rho^B$
acting on a Hilbert space $\HA \otimes \HB$,
\begin{equation} \label{eq:sumRule}
\Uni(\rho^A \otimes \rho^B) = \Uni(\rho^A) + \Uni(\rho^B) + (1-q)s \
\Uni(\rho^A)\Uni(\rho^B).
\end{equation}
Notice that  in the  cases $q=1$ or  $s=0$, one  recovers the additivity  of von
Neumann and R\'enyi entropies.

A    closed    related    concept    to    entropy    is    majorization    (see
e.g.~\cite{MarshallBook}).  Let us  consider  two density  operators $\rho$  and
$\sigma$, and  the corresponding probability vectors  $p$ and $q$  formed by the
eigenvalues  of   $\rho$  and  $\sigma$,  respectively,   sorted  in  decreasing
order. Then,  $\rho$ is majorized by  $\sigma$, denoted as  $\rho \prec \sigma$,
means that
\begin{equation} \label{eq:Majorization}
\sum_{i=1}^n p_i \, \leq \, \sum_{i=1}^n q_i \quad \mbox{for all } \ n=1,
\ldots, N-1, \quad \mbox{and } \ \sum_{i=1}^N p_i \, = \, \sum_{i=1}^N q_i,
\end{equation}
where $N=\max\left\{\rank  \rho, \rank \sigma \right\}$ and  $\rank$ denotes the
rank of  a density operator.  Notice that if  $\rank \rho \leq \rank  \sigma$ we
complete the  vector $p$ with $0$  entries to have  the same length of  $q$, and
vice versa.  This has no  impact in  the value of  unified entropies due  to the
expansibility property.

It can be  shown that $(q,s)$-entropies preserve the  majorization relation (see
e.g.~\cite{Rastegin2011, Bosyk2015}), that is,
\begin{equation} \label{eq:Schurconcavity}
\mbox{if } \ \rho \, \prec \, \sigma \quad \mbox{then} \quad \Uni(\rho) \, \geq \,
\Uni(\sigma),
\end{equation}
%
with equality if and only if $\rho$ and $\sigma$ have the same eigenvalues.
%
%
We  observe that  the reciprocal  does  not hold  in general,  which means  that
majorization is  stronger (as  an order  relation) than a  single choice  of the
entropic indexes.

Now,   using  the   Schur-concavity   it  is   straightforward   to  show   that
$(q,s)$-entropies are lower and upper bounded:
\begin{equation} \label{eq:Bounds}
0 \, \leq \, \Uni(\rho) \, \leq \, \frac{N^{(1-q)s} - 1}{(1-q)s},
\end{equation}
where the first  inequality is attained for pure states,  whereas the second one
for the maximally mixed state $\rho^*= \frac{I}{N} $.

On the other  hand, it can be  shown that the eigenvalues of  a density operator
$\rho$ are invariant under arbitrary unitary transformations $U$, in other words
$\rho$  and $U  \rho U^\dag$  have  the same  eigenvalues. Hence,  we have  that
$(q,s)$-entropies are invariant under unitary transformations
\begin{equation} \label{eq:Uinvaraince}
\Uni(\rho) = \Uni(U \rho U^\dag).
\end{equation}

Moreover, we will see in the next  subsection that the change of in entropy due
to  local  measurements   plays  a  key  role  in   order  to  quantify  quantum
correlations. Before that, we recall the  action of any bistochastic map over an
arbitrary  state.   A  bistochastic  (or completely  positive,  trace-preserving
unital) map $\E$ can be written in the Kraus form as $\E(\rho) = \sum_k E_k \rho
E_k^\dag$  with  both  sets  of  positive  operators  $\left\{E_k^\dag  E_k
\right\}$   (completely  positive)   and  $\left\{E_k   E_k^\dag  \right\}$
(unital) summing the  identity (see e.g.~\cite{BengtssonBook}). Notice that
this    map    leaves   invariant    the    maximally    mixed   state    (i.e.,
$\E(\rho^*)=\rho^*$). It can be shown that
\begin{equation} \label{eq:Bimap}
\E(\rho) \prec \rho
\end{equation}
if and  only if $\E$  is a bistochastic  map~\cite{Chefles2002},  in other  words  for bistochastic  maps the  final
state $\E(\rho)$ is more disordered  (in terms of majorization) than the initial
state $\rho$.
As   a  consequence   of~\eqref{eq:Bimap}   and  the   Schur-concavity  of   the
$(q,s)$-entropies, we have
\begin{equation} \label{eq:Bimapent}
\Uni(\E(\rho)) \geq \Uni(\rho),
\end{equation}
where the equality is attained if and only if $\E(\rho)= U \rho U^\dag$.

Hereafter, we  are only interested  in rank-one projective  measurements without
postselection, that  is, a set  orthogonal rank-one projectors  $\Pi = \{  P_i =
\ketbra{i}{i}\}$, (i.e., $P_i P_{i'} = \delta_{ii'} P_i$ and $\sum_{i=1}^N P_i=I
$) with $\{\ket{i}\}$ an orthonormal basis of $\mathcal{H}^N$. The state after a
rank-one projective measurement $\Pi$ is  equal to $\Pi(\rho) = \sum_{i=1}^N P_i
\rho P_i = \sum_{i=1}^N p_i \ketbra{i}{i}$ with $p_i = \bra{i} \rho \ket{i}$.
As projective  measurements are particular  cases of bistochastic maps,  we have
also an inequality similar to~\eqref{eq:Bimapent} for $\Pi$. Thus, we propose to
use the  difference of  $(q,s)$-entropies between the  final and  initial states
(rescaled by a factor depending of the generalized purity) as a signature of the
disturbance of the state of a system due to the measurement, that is
\begin{equation} \label{eq:Dist}
\Dist(\rho)= \frac{\Uni (\Pi(\rho)) - \Uni(\rho)}{(\Tr \rho^q)^s}.
\end{equation}
For any choice of the entropic indexes this quantity is nonnegative and vanishes
if   only  if   the  measurement   does  not   disturb  the   state  (i.e.,
$\Pi(\rho)=\rho$), which  happens when measuring in the  basis that diagonalises
$\rho$.  Notice that the rescaling  factor plays no role for von Neumann
and  R\'enyi   entropies  (additive  entropies),  on the contrary it does
for  nonadditive entropies.  In the  next  subsection, we  will clarify  the
  importance of  the rescaling $(\Tr \rho^q)^s$ when
dealing with quantum correlations measures based on nonadditive entropies.

Finally,    notice    that    two    interesting   cases    arise    from    the
definition~\eqref{eq:Dist}.  The  first  one  consists in  considering  the  von
Neumann  entropy, in  this case  the disturbance  can be  recast as  the quantum
relative  entropy (or  quantum Kullback-Leibler  divergence) between  $\rho$ and
$\Pi(\rho)$, that is
\begin{equation} \label{eq:KL}
D_{(1,s)}^\Pi(\rho) \equiv D^\Pi(\rho) = S\left(\rho \| \Pi(\rho) \right),
\end{equation}
where $S(\rho \| \sigma) = \Tr \left(\rho (\ln \rho -\ln \sigma) \right)$ is the
quantum relative entropy.
The second
one comes from evaluating~\eqref{eq:Dist} at Tsallis entropy with entropic index
equal to $2$, for which the disturbance
expresses  in  terms  of   the  Hilbert-Schmidt  distance  between  $\rho$  and
$\Pi(\rho)$ divided by the purity of $\rho$,
\begin{equation} \label{eq:HSdist}
D_{(2,1)}^\Pi(\rho)\equiv D_2^\Pi(\rho) = \frac{\| \rho - \Pi(\rho) \|^2}{\Tr
\rho^2},
\end{equation}
where $\|A\|= \sqrt{\Tr A^\dag A}$ is the Hilbert-Schmidt norm.

\subsection{Quantum  correlations from  disturbance  due to  a local  projective
  measurement}
\label{subsec:qcm}

Let  us  consider  a  bipartite   quantum  system  $AB$  with  density  operator
$\rho^{AB}$   acting   on   a   product   finite dimensional Hilbert space, $\HAB = \HA \otimes
\HB$,  where  $N^{AB} = N^A  N^B$.  Following~\cite{Luo2008}, we consider the
local rank-one projective measurements (without postselection), $\Pi^A = \{P^A_i
\otimes I^B \}$, $\Pi^B= \{I^A \otimes P^B_j \}$ and $\Pi^{AB}= \{ P^A_i \otimes
P^B_j \}$, where $\left\{P^A_i \right \}  $ and $\left\{P^B_j \right \}$ are set
of orthogonal  rank-one projectors  that sum to  the identity, $I^A$  and $I^B$,
respectively.  Then, the resulting states after these measurements are
\begin{align}
\Pi^A(\rho^{AB}) & = \sum_{i} P^A_i \otimes I^B \ \rho^{AB} P^A_i \otimes I^B =
\sum_{i} p^A_i P^A_i \otimes \rho^{B\mid i}, \label{eq:CQS}\\[2mm]
\Pi^B(\rho^{AB}) & = \sum_{j} I^A \otimes P^B_j \ \rho^{AB} I^A \otimes P^B_j =
\sum_{j} p^B_j \rho^{A\mid j} \otimes P^B_j, \label{eq:QCS}\\[2mm]
\Pi^{AB}(\rho^{AB}) & = \Pi^A\circ\Pi^B(\rho^{AB}) = \Pi^B\circ\Pi^A(\rho^{AB})
\nonumber \\[2mm]
& = \sum_{ij} P^A_i \otimes P^B_j \ \rho^{AB} P^A_i \otimes P^B_j = \sum_{ij}
p^{AB}_{ij} P^A_i \otimes P^B_j, \label{eq:CCS}
\end{align}
where   $\ \rho^{B\mid   i} = \frac{\Tr_{A}\left(P^A_i \otimes I^B \rho^{AB}\right)}{p^A_i}\ $  with
$\ p^A_i  =  \Tr\left(  P^A_i  \otimes  I^B  \rho^{AB}\right)$,
$\ \rho^{A\mid   j} = \frac{\Tr_{B}\left(I^A   \otimes   P^B_j   \rho^{AB} \right)}{p^B_j}\ $
      with $\ p^B_j = \Tr\left( I^A  \otimes P^B_j \rho^{AB}\right)\ $ and $\ p^{AB}_{ij} = \Tr\left( P^A_i \otimes P^B_j \rho^{AB}\right)$.
According  to~\cite{Luo2008}, these  states are  called  classical-quantum (CQ),
quantum-classical  (QC)  and  classical-classical  (CC) correlated  states  with
respect   to   the  local   measurements   $\Pi^A$,   $\Pi^B$  and   $\Pi^{AB}$,
respectively. A state is said CQ correlated if there is a local projective measurement over
$A$   that  does   not  disturb   it,   i.e.,  $\Pi^A(\rho^{AB})=\rho^{AB}$
(analogously for  QC and CC correlated  states). All these  states are separable
(i.e.,  nonentangled),   as  they  are  convex   combinations  of  product
states~\cite{Werner1989},   although  not   all   separable  states   are  of   the
forms~\eqref{eq:CQS}--\eqref{eq:CCS}. Moreover,  the sets  formed by all  CQ, QC
and CC correlated states, denoted as $\Omega^A$, $\Omega^{B}$ and $\Omega^{AB}$,
respectively, are not convex in contrast  to the set of separable states. Notice
that  $\Omega^{A}$  and $\Omega^{B}$  are  the  sets  of zero  quantum
discord   states  with  respect   to  $\HA$   and  $\HB$
  respectively,  and $\Omega^{AB}  =  \Omega^A \cap  \Omega^B$~\cite{Dakic2010,
  Datta2011}. In the  sequel, for sake of brevity, we will  use $L$ to denote either  $A$ or $B$,
  and $K$ to denote $A$, $B$ or $AB$.

Now, we  can use~\eqref{eq:Dist}  to quantify the  disturbance due to  the local
projective measurement $\Pi^{K}$,
\begin{equation} \label{eq:QCk}
\MDk = \frac{\Uni \left(\Pi^{K}(\rho^{AB})\right) - \Uni(\rho^{AB})}{\Pty}.
\end{equation}
We   denote  $D_{(q,s)}^{\Pi^{L}}$   as   unilocal  disturbances,   whereas
$D_{(q,s)}^{\Pi^{AB}}$ as bilocal disturbances.

In order to obtain  a measurement-independent signature of quantum correlations,
one takes the  minimum of the disturbances~\eqref{eq:QCk} over  the set of local
measurements, that is
\begin{equation} \label{eq:mQCk}
\mMDk = \min_{{\Pi^K}} \MDk.
\end{equation}
The  following properties  justify our  proposal~\eqref{eq:mQCk} as  measures of
quantum correlations:
\begin{enumerate}[(i)]
 \item nonnegativity: $\mMDk \geq 0$ with equality if and only if $\rho^{AB} \in \Omega^K$. Accordingly,
$D_{(q,s)}^{L}$ are semiquantum  correlations measures (with respect to
  $\mathcal{H}^{N^L}$), whereas $D_{(q,s)}^{AB}$ are total quantum correlation measures;
\item  invariance  under local  unitary  operators:  $D_{(q,s)}^K(U\otimes V  \,
  \rho^{AB} \, U^\dagger\otimes V^\dagger) =  \mMDk$, where $U$ and $V$ are
  a unitary operations over $A$ and $B$ respectively; and
\item invariance  when an uncorrelated  ancilla is appended to  the system:
  $D_{(q,s)}^{K}(\rho^{AB}  \otimes   \rho^C)  =  D_{(q,s)}^{K}(\rho^{AB})$  for
  bipartitions  $A|BC$  or  $B|AC$  (for  the  bipartition  $AB|C$  the  quantum
  correlations measures naturally vanish).
\end{enumerate}
The first property is a  direct consequence of majorization relation between the
states after  and before local projective  measurements.  The second  one can be
proved from the definition of our measure, Eq.~\eqref{eq:mQCk}, noting that
$\Pi^K(U\otimes  V \,  \rho^{AB} \,  U^\dagger\otimes V^\dagger))=U\otimes  V \,
\tilde{\Pi}^K(\rho^{AB})     \,    U^\dagger\otimes     V^\dagger     $,    with
$\tilde{\Pi}^K=U^\dagger\otimes V^\dagger \, \Pi^K \, U\otimes V$, and recalling
the invariance  of $(q,s)$-entropies  under unitary transformations.   The third
property is more subtle and it  is related to the sum rule~\eqref{eq:sumRule} of
the  $(q,s)$-entropies. Indeed,  the generalized  purity factor  $\Pty$  plays a
crucial  role to fulfill  this property  in the  case of  nonadditive entropies,
without affecting the  complexity of computability of the  measures. In general,
this property has not been taking  into account in the literature of nonadditive
entropic  measures  of  quantum  correlations. For  instance,  entropic  quantum
correlations  measures based  on the  difference of  trace form  entropies\footnote{Notice that $(q,s)$-entropies reduce  to a trace  form only if  $s=1$ (Tsallis entropies).}, i.e., $S_\phi(\rho) = \Tr
\phi(\rho)$    with     $\phi$    concave    and    $\phi(0)=0$~\cite{Wehrl1978},
have been  dealt in
Refs.~\cite{Rossignoli2010,  Costa2013}. However,  these  measures are  not
invariant when  an uncorrelated ancilla  is appended to the  system, except
for the von Neumann case.  This  is direct consequence of nonadditivity of trace
form entropies.   For a more  general discussion about necessary  and reasonable
conditions of quantum correlations measures, see~\cite{Brodutch2012}.  Moreover,
our  semiquantum correlations  measures can  be  also interpreted  as a  quantum
deviation  from  the  Bayes  rule  in  a  way  similar  to  that  discussed
in~\cite{Costa2013}.

We remark  that our quantum  correlations measures include some  important cases
already   discussed   in   the   literature.   The   first   one   consists   in
evaluating~\eqref{eq:mQCk} for the von Neumann entropy. In this case we reobtain
the so-called  information deficit~\cite{Horodecki2005}, which  can be rewritten
in    terms    of    the    minimal    relative   entropy    over    the    sets
$\Omega^K$~\cite{Modi2010},
\begin{equation}
D^{K}(\rho^{AB}) = \min_{\Pi^K} S\left(\rho^{AB} \| \Pi^{K}(\rho^{AB}) \right)
 = \min_{\chi^{AB} \in \Omega^{K}} S\left(\rho^{AB} \| \chi^{AB} \right).
\end{equation}
The second  one arises when  evaluating~\eqref{eq:mQCk} for the  Tsallis entropy
with  entropic  index  equal  to  $2$.  This case  is  close  to  the  geometric
discord~\cite{Dakic2010},
\begin{equation}
D_G^{K}(\rho^{AB}) = \min_{\chi^{AB} \in \Omega^{K}} \| \rho^{AB} - \chi^{AB} \|^2.
\end{equation}
Indeed,  using  the   expression  of  $D_G^K$  in  terms   of  local  projective
measurements given in~\cite{Luo2010}, we obtain
\begin{equation}
D_2^K(\rho^{AB}) = \frac{\min_{\Pi^{K}} \| \rho^{AB} - \Pi^{K}(\rho^{AB})
\|^2}{\Tr (\rho^{AB})^2} = \frac{D_G^K(\rho^{AB})}{\Tr (\rho^{AB})^2}.
\end{equation}
Notice that $D_G^{K}$ is not  invariant when an uncorrelated ancilla is
  appended to the system~\cite{Piani2012}.
The  purity  rescaled factor  solves  this issue~\cite{Tufarelli2013},  although
there is not the unique  way to do it (see e.g~\cite{Tufarelli2013, Chang2013}).
Finally, notice that  in the case of R\'enyi  entropies, which has recently
been  introduce  in~\cite{Canosa2015},  our  measure  fulfills  the  desired
invariance property when appending an uncorrelated ancilla to the system.

\subsection{Lower bound and its relation with entanglement}
\label{subsec:lowerbound}

First,   let   us   note   that    since   QC,   CQ   and   CC   correlated
states~\eqref{eq:CQS}--\eqref{eq:CCS}  are  separable,  they  fulfill  some
general entropic separability inequalities (see e.g.~\cite{Bosyk2015}),
\begin{equation} \label{eq:ent_ineq}
\Uni(\Pi^K(\rho^{AB})) \geq \max \left\{ \Uni(\Tr_A \Pi^K(\rho^{AB})),
\Uni(\Tr_B\Pi^K(\rho^{AB})) \right\}.
\end{equation}
On the other hand, the corresponding final reduced states are
\begin{align}
\Tr_A \Pi^A(\rho^{AB}) = \rho^B \ &\mbox{and} \ \Tr_B \Pi^A(\rho^{AB}) = \Tr_B
\Pi^{AB}(\rho^{AB}) = \sum_{i} p_i^A P^A_i
=\rho^{A}_{\mathrm{diag}}, \label{eq:reducedCQS}\\
\Tr_B \Pi^B(\rho^{AB}) = \rho^A \ &\mbox{and} \ \Tr_A \Pi^B(\rho^{AB}) = \Tr_A
\Pi^{AB}(\rho^{AB}) = \sum_{j} p^B_j P^B_j =
\rho^{B}_{\mathrm{diag}}. \label{eq:reducedQCS}
\end{align}
where $\rho^{L}_{\mathrm{diag}}$  denotes the diagonal of  $\rho^{L}$ in the
  basis     underlying    by $\{P_i^{L}\}$.      Since
  $\rho^{L}_{\mathrm{diag}} \prec \rho^{L}$~\cite{BengtssonBook} and due to the
Schur-concavity of the $(q,s)$-entropies, inequality~\eqref{eq:ent_ineq} reduces
to
\begin{equation} \label{eq:ent_ineq_simpler}
\Uni(\Pi^K(\rho^{AB})) \geq \max \left\{ \Uni(\rho^{A}), \Uni(\rho^{B})\right\}.
\end{equation}
Thus,  plugging~\eqref{eq:ent_ineq_simpler} into~\eqref{eq:QCk} to
  lowerbound $\MDk$ and taking the minimum,
we obtain that the quantum correlations measures are lower bounded, as follows
\begin{equation} \label{eq:lower_bound_ent}
\mMDk \geq \max \left\{ \frac{\Uni(\rho^{A})-\Uni(\rho^{AB})}{\Pty} ,
\frac{\Uni(\rho^{B})-\Uni(\rho^{AB}) }{\Pty} \right\}.
\end{equation}
Notice  that this  lower bound  could be  nontrivial only  for  entangled sates;
indeed,  the  right  hand  side of~\eqref{eq:lower_bound_ent}  is  negative  for
separable states.   A similar result  has already been  obtained in the  case of
trace form entropies~\cite{Rossignoli2010}.

Now, let  us consider a pure state  $\rho^{AB} = \ketbra{\Psi^{AB}}{\Psi^{AB}}$.
Let us suppose that
\begin{equation} \label{eq:Schmidt}
\ket{\psi^{AB}}=\sum_{k=1}^{n}{\sqrt{\lambda_k}\ket{k^A}\otimes\ket{k^B}}
\end{equation}
is the Schmidt decomposition  of $\ket{\psi^{AB}}$ ($n \le \min(\{N^A,N^B\}$
  and $\{\ket{k^L}\}$ are  a orthonormal set).  Thus, it  can be
shown    that     the    reduced     states     $\rho^A    =     \Tr_B
  \ketbra{\Psi^{AB}}{\Psi^{AB}}$        and        $\rho^B        =        \Tr_A
  \ketbra{\Psi^{AB}}{\Psi^{AB}}$   have  the  same   unified  entropy
and,  as  a  consequence, the  lower  bound~\eqref{eq:lower_bound_ent}
reduces  to $\Uni(\rho^{A})= \Uni(\rho^{B})$  for pure  states $\rho^{AB}$.
Moreover,this bound  is saturated when the  local measurements are taking  in the Schmidt
basis.  After  these measurements, i.e.,  choosing the local  projectors as
$P^L_k = \ketbra{k^L}{k^L}$  (completed   to   obtain   $N^L$   projector),   the  state   is   given   by
$\Pi^K(\rho^{AB})  = \sum_{k}  \lambda_{k}  P^A_k \otimes  P^B_k$, with  unified
entropies $\Uni(\Pi^K(\rho^{AB}))  = \Uni(\rho^{A})= \Uni(\rho^{B})$. Therefore,
we  obtain that  for  pure  states the  entropic  quantum correlations  measures
becomes a generalization of the entanglement entropy,
\begin{equation} \label{eq:equality}
\mMDk = \Uni(\rho^A) = \Uni(\rho^B),
\end{equation}
which for the von Neumann entropy reduces to the standard one~\cite{Bennett1996}.

\subsection{Relationships between total and semiquantum correlations}
\label{subsec:total_semi_qcm}
It  is  possible  to  find  some interesting  relationships  between  total  and
semiquantum correlations  when bilocal  disturbances, $\MDAB$, are  rewritten in
terms of unilocal disturbances,
\begin{align}
\MDAB = \MDA + \Pc{\Pi^A} \MDBc{\Pi^A(\rho^{AB})} \label{eq:bilocal_a} \,, \\
\MDAB = \MDB + \Pc{\Pi^B} \MDAc{\Pi^B(\rho^{AB})} \label{eq:bilocal_b} \,,
\end{align}
where  $\Pc{\Pi}   =  \left(  \frac{\Tr(\Pi(\rho^{AB}))^q}{\Tr(\rho^{AB})^q}
  \right)^s$  (for sake of brevity, we omit the dependence  of this factor on the state).
  This quantity, $\Pc{\Pi}$, is  nonnegative but  it can take
values below or above $1$, depending on the value of the entropic parameter $q$.
As  $\Pi(\rho)\prec\rho$,  we have  that  $\Pi(\rho)^q\prec\rho^q$ if  $q\geq1$,
whereas, $\rho \prec \Pi(\rho)$ holds if $0\leq q<1$. Thus, $\Pc{\Pi}\in(0,1]$
if  $q\geq1$, else  $\Pc{\Pi}\geq1$. In  particular, for  Rényi  entropies the
factor is always equal to $1$.

Now, let us consider two possible measurement scenarios:
\begin{itemize}
\item $\Pi_0^{AB}=\Pi_0^A \circ \Pi_0^B$ is a bilocal measurement that minimizes
  the       total       quantum       correlation      measure,       i.e.,
  ${\MDABcc{0}{\rho^{AB}}=\mMDAB}$,
\item ${\Pi_1^{AB}=\Pi_1^A\circ\Pi_1^B}$, where ${\Pi_1^{L}}$, optimize the
  unilocal   disturbances,   i.e.,   $D_{(q,s)}^{\Pi_{1}^{L}}   (\rho^{AB})=
    D_{(q,s)}^{L}(\rho^{AB})$.
\end{itemize}
Applying  Eqs.~\eqref{eq:bilocal_a}--\eqref{eq:bilocal_b} to both  scenarios, we
obtain
\begin{equation} \label{eq:bilocal_opt0}
\mMDAB = \MDAcc{0}{\rho^{AB}} + \Pc{\Pi_0^A} \MDBcc{0}{\Pi_0^A(\rho^{AB})}%
= \MDBcc{0}{\rho^{AB}} + \Pc{\Pi_0^B} \MDAcc{0}{\Pi_0^B(\rho^{AB})} \,,
\end{equation}
and
\begin{equation} \label{eq:bilocal_opt1}
\MDABcc{1}{\rho^{AB}} = \mMDA + \Pc{\Pi_1^A} \MDBcc{1}{\Pi_1^A(\rho^{AB})}%
 = \mMDB + \Pc{\Pi_1^B} \MDAcc{1}{\Pi_1^B(\rho^{AB})} \,.
\end{equation}
Using that  ${\mMDAB \leq  \MDABcc{1}{\rho^{AB}}}$ (and the  analogous relations
for             the             unilocal            disturbances)             on
Eqs.~\eqref{eq:bilocal_opt0}--\eqref{eq:bilocal_opt1}  respectively, it can
been shown that ${\mMDAB}$ is lower and upper bounded as follows,
\begin{align}
\mMDAB \geq \max\{ \mMDA + \Pc{\Pi_0^A}\MDBcc{0}{\Pi_0^A(\rho^{AB})} , \mMDB +
\Pc{\Pi_0^B}\MDAcc{0}{\Pi_0^B(\rho^{AB})} \} \label{eq:lower_bound}\,,\\
\mMDAB \leq \min\{ \mMDA + \Pc{\Pi_1^A}\MDBcc{1}{\Pi_1^A(\rho^{AB})} , \mMDB +
\Pc{\Pi_1^B}\MDAcc{1}{\Pi_1^B(\rho^{AB})} \} \label{eq:upper_bound}\,.
\end{align}
In    particular,   given    that   the    nonoptimal    unilocal   disturbances
in~\eqref{eq:lower_bound}  are  nonnegative,  we  naturally  obtain  that  total
quantum correlation are greater than or equal to the semiquantum ones,
\begin{equation} \label{eq:lower_bound_b}
\mMDAB \geq \max\{ \mMDA , \mMDB \}.
\end{equation}
This  result  can  be  also  obtained  more  directly  from  the  fact  that
  $\Uni(\Pi^{AB}_0(\rho^{AB}))    \geq   \Uni(\Pi^{L}_1(\rho^{AB}))$.    Notice
that~\eqref{eq:lower_bound_b}  is  in accordance  with  the inclusion  relations
among the  sets of CQ,  QC and CC  correlated states, i.e.,  $\Omega^{AB} =
  \Omega^A \cap \Omega^B \subset \Omega^{L}$.

Moreover, noting that  $2 \, \MDAB \ge  \mMDAB + \MDAB \ge 2  \, \mMDAB$ we
can  deduce from  Eqs.~\eqref{eq:bilocal_opt0}--\eqref{eq:bilocal_opt1} the
following inequality for the sum of semiquantum correlations:
\begin{equation} \label{eq:DADB}
\mMDAB + \Delta_0 \geq \mMDA + \mMDB \geq \mMDAB + \Delta_1 \,,
\end{equation}
where  we   defined  the   quantities  ${\Delta_i  :=   \MDABcc{i}{\rho^{AB}}  -
  \Pc{\Pi_i^B}        \MDAcc{i}{\Pi_i^B(\rho^{AB})}        -        \Pc{\Pi_i^A}
  \MDBcc{i}{\Pi_i^A(\rho^{AB})}}$, with ${i=0,1}$.

Notice that for CQ and  QC correlated states, one has ${\Delta_1=0}$, $\Pi_1^L$
  being defined by
the set $\{P^L_i\}$ so that it does not disturb the joint state.
Finally,
notice  that  for  CC   correlated  states,  all  quantities  in~\eqref{eq:DADB}
vanish.       Therefore,       from       these      observations       together
with~\eqref{eq:lower_bound_b}, we obtain
\begin{itemize}
 \item if $\mMDA = 0$, then $\mMDAB=\mMDB$,
 \item if $\mMDB = 0$, then $\mMDAB=\mMDA$,
 \item if $\mMDAB = 0$, then $\mMDA=\mMDB=0$.
\end{itemize}
%
Furthermore, a triangle-like inequality between total and semiquantum correlations,
\begin{equation} \label{eq:triangle_ineq}
 \mMDA + \mMDB \geq \mMDAB,
\end{equation}
is trivially satisfied for CQ, QC and CC correlated states.
The  validity of  the triangle-like  inequality~\eqref{eq:triangle_ineq}  in the
general   case    relies   on   the   sign   of    $\Delta_1$.   If   ${\Delta_1
  \geq0\,\forall\rho^{AB}}$, the inequality is  generally true. On the contrary,
if  ${\Delta_1 <0}$ for  some $\rho^{AB}$  then it  could be  the case  that the
inequality does not hold for those states.

Although  the most  general conditions  for  the validity  of the  triangle-like
inequality~\eqref{eq:triangle_ineq}  are  hard  to  analyze,  we  can  link  the
validity of~\eqref{eq:triangle_ineq} with a kind of local contractivity property
of the unilocal disturbances. Specifically, let us assume as valid the following
inequalities:
\begin{align}
\Pc{\Pi_j^B}\MDAcc{i}{\Pi_j^B(\rho^{AB})} &\leq
\MDAcc{i}{\rho^{AB}}, \label{eq:contractivity_a} \\
\Pc{\Pi_j^A}\MDBcc{i}{\Pi_j^A(\rho^{AB})} &\leq
\MDBcc{i}{\rho^{AB}}\label{eq:contractivity_b}.
\end{align}
Then, we have
\begin{align}
\Pc{\Pi_1^B}\MDAcc{1}{\Pi_1^B(\rho^{AB})} &\leq \MDAcc{1}{\rho^{AB}} = \mMDA, \\
\Pc{\Pi_1^A}\MDBcc{1}{\Pi_1^A(\rho^{AB})} &\leq \MDBcc{1}{\rho^{AB}} = \mMDB,
\end{align}
and, replacing any of these relations in~\eqref{eq:bilocal_opt1}, we obtain
\begin{equation}
\MDABcc{1}{\rho^{AB}} \leq \mMDA + \mMDB.
\end{equation}
Finally,  recalling  that  ${\mMDAB\leq\MDABcc{1}{\rho^{AB}}}$, it  follows  the
triangle-like inequality~\eqref{eq:triangle_ineq}.

\begin{figure}
\centering
\includegraphics[width=.95\textwidth]{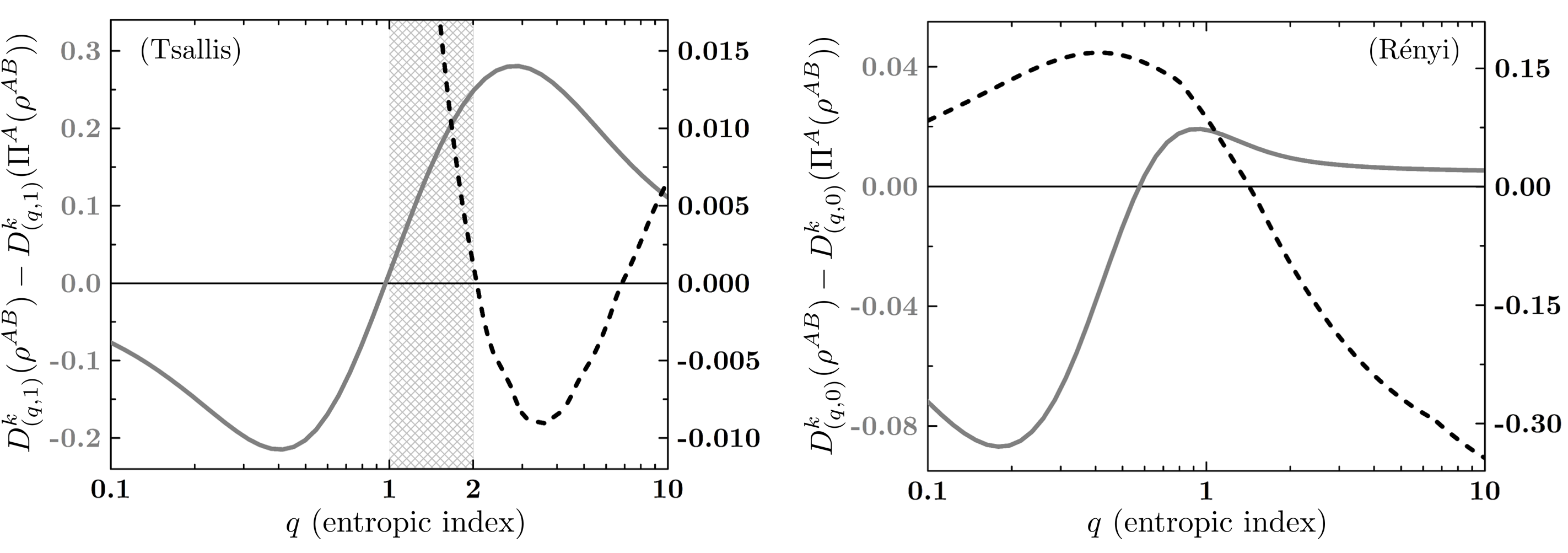}
\caption{Minimal     differences    between    $\MDA$     and    $\Pc{\Pi^A}
    \MDAc{\Pi^B(\rho^{AB})}$  computed  for   $10^3$  random  local  projective
  measurements $\Pi^{A(B)}$, using Tsallis entropies (left figure) and Rényi
  entropies  (right figure).  Each line  corresponds to  a  random two-qubit
  state.   Notice that  a wide  range of  values of  the parametric  index, $q$,
  yields  negative values  for these  differences, implying  a violation  of the
  contractivity    property   under    local   projective    measurements   (see
  relations~\eqref{eq:contractivity_a}--\eqref{eq:contractivity_b}  and text for
  details). For $q=1$ both measures converge to the von Neumann-based one, which
  fulfills  the  contractivity  property.   The  same happens for  Tsallis  with  $q=2$,
  corresponding to  the Hilbert-Schmidt distance. Interestingly,  in the Tsallis
  case  we  have  been  unable   to  find  a  counterexample  to  the  mentioned
  contractivity for $q\in(1,2)$ (shaded region of the left figure).}
  \label{fig:contractivity}
\end{figure}

Thus, we are able to link  the validity of the triangle-like inequality, for all
states  and  any entropic  indexes,  with  the  assumption of  contractivity  of
unilocal disturbances  under local projective  measurements. In the case  of von
Neumman                                                                  entropy,
inequalities~\eqref{eq:contractivity_a}--\eqref{eq:contractivity_b}           are
particular  cases of  the contractivity  of the  quantum relative  entropy under
trace-preserving  completely positive  maps~\cite{Lindblad1975}.  Otherwise, for
Tsallis          entropy         of         entropic          index         $2$,
inequalities~\eqref{eq:contractivity_a}--\eqref{eq:contractivity_b}           are
particular  cases of  the contractivity  of the  Hilbert-Schmidt  distance under
projective   measurements~\cite{Tamir2015}.  Therefore,   in   both  cases   the
triangle-like inequality is  satisfied (notice that for the  latter, this result
has been  proved in  alternative way~\cite{Zhao2013}). Unfortunately,  the local
contractivity is  not valid  for general entropic  functionals. Indeed,  we show
that is  the case  for a wide  range of  the entropic index  of the  R\'enyi and
Tsallis entropies in Fig.~\ref{fig:contractivity}.

\section{Examples}
\label{sec:examples}

\subsection{Mixtures of a pure state and the maximally mixed one}

An interesting  example where the  computations can be carried  out analytically
involves  the family  of \textit{pseudopure}  states,  given by  mixtures of  an
arbitrary   pure   state,    $\ket{\psi^{AB}} \in \HA \otimes \HB$, with the maximally mixed state, yielding
\begin{equation} \label{eq:pseudopure}
\rho^{AB}_p = (1-p) \frac{I^{AB}}{N^{AB}} + p \ketbra{\psi^{AB}}{\psi^{AB}},
\end{equation}
with  $0\leq p\leq1$  (remind that  $N^{AB} =  N^A N^B$).  The  spectrum of
$\rho^{AB}$ is given by  the eigenvalue $(1-p)/N^{AB}+p$, with multiplicity
$1$,    and    the    eigenvalues   $(1-p)/N^{AB}$,    with    multiplicity
$N^{AB}-1$.  The  measurements that  optimize  both  the  unilocal and  the
bilocal quantifiers  are unique  (do not  depend on the  entropic form)  and are
given    by     the    local    Schmidt     basis~\cite{Rossignoli2010}.    This
entropic-independent measurement  fact is not a universal  property, but depends
on the particular states. In this  case, measuring in the Schmidt basis yields a
final  spectrum that is  majorized by  any other  spectrum corresponding  to any
other measurement, implying the entropic-independent optimization.
After   the   measurement,   the   spectrum   is   given   by   the   eigenvalue
$(1-p)/N^{AB}$,  with  multiplicity  $N^{AB}-n$, and  the  eigenvalues
$(1-p)/N^{AB} + p  \lambda_k$ with $1\leq k\leq n$, where  $n$ is the Schmidt
number and $\lambda_k$ the square of Schmidt coefficients~\eqref{eq:Schmidt}. Using Eq.~\eqref{eq:mQCk}, we obtain
\begin{equation} \label{eq:pseudopure3}
D^K_{(q,s)}(\rho^{AB}_p) = \frac{1}{(1-q)s} \left[ \left(\frac{(N^{AB} -
n)(1-p)^q + \sum_{k=1}^{n}{[1+(N^{AB} \lambda_k - 1) p]^q}}{( N^{AB} -
1)(1-p)^q + [1 + ( N^{AB} - 1) p]^q}\right)^s - 1 \right]
\end{equation}
for the generalized quantum correlations  of pseudopure states. It is remarkable
that, in  this particular  case and  given the collapse  of the  semiquantum and
total  quantifiers, the triangle-like  inequality~\eqref{eq:triangle_ineq} holds
for the most general $(q,s)$-entropic forms.

In  particular, when  $\ket{\psi^{AB}}$  is a  maximally  entangled state,  with
$N^A=N^B=N$,  states $\rho^{AB}_p$  constitutes  a family  of isotropic  states,
$\rho^I_p$. In  that case, $\forall  k, \quad \lambda_k=N^{-1}$,  $n=N$, and
  the generalized quantum correlations are
\begin{equation} \label{eq:isotropic}
D^K_{(q,s)}(\rho^I_p) = \frac{1}{(1-q)s} \left[ \left( \frac{(N^2-N)(1-p)^q + N
[ 1 + (N-1) p ]^q}{( N^2 - 1)(1 - p)^q + [1 + (N^2-1) p]^q} \right)^s - 1
\right].
\end{equation}
Specializing this for Tsallis and R\'enyi entropies one obtains, respectively,
\begin{align}
D^K_{(q,1)}(\rho^I_p) &= \frac{1}{1-q} \frac{1}{N^{2q}} \left[ N (1-p+Np)^q -
(1-p+N^2p)^q (N-1) (1-p)^q \right], \\
D^K_{(q,0)}(\rho^I_p) &= \frac{1}{1-q} \ln\left[ \frac{N (1-p+Np)^q + (N^2-N)
(1-p)^q}{(1-p+N^2p)^q + (N^2-1) (1-p)^q} \right].
\end{align}

\subsection{Werner and isotropic states}

Although  isotropic states  are particular  cases  of Eq.~\eqref{eq:pseudopure},
i.e., mixtures of a pure state and  the maximally mixed one, we aim to show
that  both isotropic~\cite{Horodecki1999}  and  Werner states~\cite{Werner1989},
due to their symmetries, are  independent of the local measurements performed. A
Werner  state is  a $N  \times N$  dimensional bipartite  quantum state  that is
invariant under  local unitary transformations of  the form $U  \otimes U$, with
$U$ an arbitrary unitary acting on $N$ dimensional systems, that is, $\rho^W = U
\otimes  U \rho U^\dagger  \otimes U^\dagger$.  On the  other hand,  an $N\times
N$-dimensional isotropic  state is invariant under arbitrary  local unitaries of
the form $U \otimes U^\ast$, that  is, $\rho^I = U \otimes U^\ast \rho U^\dagger
\otimes (U^\ast)^\dagger$. They can be parametrized, respectively, as
\begin{equation} \label{eq:werner_states}
\rho^W_x = \frac{N-x}{N^3-N} I + \frac{Nx-1}{N^3-N} F,
\end{equation}
with $F=\sum_{ij}\ketbra{ij}{ji}$, $1\leq i,j\leq N$, $x\in[-1,1]$, and
\begin{equation} \label{eq:isotropic_states}
\rho^I_y = \frac{1-y}{N^2-1} I + \frac{N^2y-1}{N^2-1} \ketbra{\psi^+}{\psi^+},
\end{equation}
with  $\ket{\psi^+} =  \frac{1}{\sqrt{N}} \sum_{i=1}^N{\ket{ii}}$  and $y  \in [
\frac{1}{N^2} , 1 ]$. Notice that both definitions of isotropic states --the one
derived    from    Eq.~\eqref{eq:pseudopure}    and    the    one    given    by
Eq.~\eqref{eq:isotropic_states}--  coincide under  the  identification $p  =
  \frac{N^2y-1}{N^2-1}\ $ and $\ \ket{\psi^{AB}} = \ket{\psi^+}$.

To  see  that any  local  measurement yields  the  same  disturbance over  these
families  of  states,  let  us   consider  $\Pi^A_1$  as  the  optimal  unilocal
measurement  over $A$.  Any  other local  measurement is  achieved by  a unitary
transformation  over $\Pi^A_1$ as  $\Pi^A_V =  V \otimes  I^B \,  \Pi^A_1 \,
  V^\dagger \otimes I^B$,  with $V$ an arbitrary unitary  over $A$. Then, using
the  invariance  properties  of  Werner  states,  the  action  of  $\Pi^A_V$  is
$\Pi^A_V(\rho^W)   =   V   \otimes   V   \,   \Pi^A_1   \,   V^\dagger   \otimes
V^\dagger$. Analogous  results holds for isotropic states  and measurements over
$B$.   Invoking the  unitary invariance  of $(q,s)$-entropies  one has  that the
minimum in~\eqref{eq:mQCk} is attained  for any local projective measurement. To
prove  that  nothing  changes  when  considering  bilocal  measurements,  it  is
sufficient to  observe that after any  local measurement the state  becomes a CC
correlated state. Thus, given that the total disturbance can be computed via the
partial disturbances  (see Eqs.~\eqref{eq:bilocal_a}--\eqref{eq:bilocal_b}), the
total quantum correlations are equal to the semiquantum ones.

In order  to find  an explicit  formula of the  generalized correlations,  it is
easier to measure on  the standard basis (the ones used to  define $F$ in Werner
states and $\ket{\psi^+}$ in isotropic states), readily obtaining
\begin{equation} \label{eq:werner_qc}
D^K_{(q,s)}(\rho^W_x) = \frac{1}{(1-q)s} \left[ \left( \frac{2[(N-1)^q (x+1)^q +
(N-1) (N-x)^q]}{2 (N-1)^q (x+1)^q + (N-1) [(N - x + \frac{1}{2} N x -
\frac{1}{2})^q + (N - x - \frac{1}{2} N x + \frac{1}{2})^q]} \right)^s - 1
\right]
\end{equation}
and
\begin{equation} \label{eq:isotropic_qc}
D^K_{(q,s)}(\rho^I_y) = \frac{1}{(1-q)s} \left[ \left( \frac{N [(N-1) (1-y)^q +
(1 - y + N y - \frac{1}{N})^q]}{(N^2 - 1)^q y^q + (N^2 - 1) (1 - y)^q} \right)^s
- 1 \right]
\end{equation}

Again, it is interesting to observe  that these families of states are among the
ones  that satisfy  the triangle-like  inequality (Eq.~\eqref{eq:triangle_ineq})
for any $(q,s)$-entropy.

\section{Concluding remarks}
\label{sec:conclusion}

In this work  we address the problem of  quantifying quantum correlations beyond
discord. Specifically, following~\cite{Luo2008},  we obtain entropic measures of
bipartite quantum  correlations by  quantifying the system's  states disturbance
under local measurements. Our measures  are based on very general entropic forms
given by the $(q,s)$-entropies. As a consequence, we obtain quantum correlations
measures,  which include  as  particular cases  or  are close  to several  other
measures    previously   discussed   in    the   literature~\cite{Rajagopal2002,
  Horodecki2005,  Luo2008, Modi2010, Dakic2010,  Rossignoli2010, Rossignoli2011,
  Canosa2015}.   Our   main  contribution   is  to  propose   such  quantum
correlations measures based on  quantum unified $(q,s)$-entropies that are:
(i)  nonnegative and  vanishes only  for QC,  CQ and  CC correlated  sates, (ii)
invariant under local unitary operators,  and (iii) invariant under the addition
of an uncorrelated  ancilla. Regarding with the last property,  we show that for
$q   \not\rightarrow  1$   or  $s   \not\rightarrow   0$,  that   is  when   the
$(q,s)$-entropies are  nonadditive, it is necessary to  rescale the disturbances
by a generalized purity factor in order to avoid undesirable effects of previous
entropic   based   correlation  measures~\cite{Rossignoli2010,   Rossignoli2011,
  Costa2013}.

Moreover, we  distinguish between  total and  semiquantum correlations,  and  we naturally
obtain that the former are greater than  the latter. In addition, we show that a
triangle-like inequality is fulfilled for  certain families of sates, namely QC,
CQ and  QQ correlated states, as well  as, Werner and Isotropic  states, for any
entropic measures. In  the general case, we only proof this  for the von Neumann
and  Tsallis  with  entropic  index   of  order  $2$,  which  follows  from  the
contractivity  property  under  a  projective measurement  of  quantum  relative
entropy  and  Hilbert  Schmidt  distance,  respectively.  We  provide  numerical
counterexamples where the local  contractivity property of unilocal disturbances
fails in  a wide range of the  entropic index of R\'enyi  and Tsallis entropies,
but  it remains  open if  the triangle-like  inequality is  fulfilled  for other
entropic measures.

Finally, we provide analytical expressions of the entropic correlations measures for pseudopure, Werner and isotropic states. For these families of states, the optimal measurement of unilocal and bilocal disturbances are independent of the entropic form.

\section*{Acknowledgments}

GMB, GB,  MP and PWL  acknowledge CONICET and  UNLP (Argentina), and MP  and PWL
also acknowledge SECyT-UNC (Argentina) for  financial support. SZ is grateful to
the University of Grenoble-Alpes (France) for the AGIR financial support.


\end{document}